%% file: TAUP15Proceedings_blenert_v3.tex
\documentclass[a4paper]{jpconf}
\usepackage{graphicx}

\usepackage{units}
\usepackage{color}

\begin{document}
\include{commands}
\title{Background Rejection of n$^+$ Surface Events in GERDA Phase II}

\author{Bj\"{o}rn Lehnert\footnote{for the GERDA Collaboration}}

\address{Institut f\"{u}r Kern- und Teilchenphysik\\ Technische Universit\"{a}t Dresden, Germany}

\ead{bjoernlehnert@gmail.com}

\begin{abstract}
The \textsc{Gerda} experiment searches for neutrinoless double beta ($0\nu\beta\beta$) decay in $^{76}$Ge using an array of high purity germanium (HPGe) detectors immersed in liquid argon (LAr). Phase II of the experiment uses 30 new broad energy germanium (BEGe) detectors with superior pulse shape discrimination capabilities compared to the previously used semi-coaxial detector design.
By far the largest background component for BEGe detectors in \textsc{Gerda} are n$^+$-surface events from $^{42}$K $\beta$ decays which are intrinsic in LAr. The $\beta$ particles with up to 3.5~MeV can traverse the 0.5 to 0.9~mm thick electrode and deposit energy within the region of interest for the $0\nu\beta\beta$ decay. However, those events have particular pulse shape features allowing for a strong discrimination.
The understanding and simulation of this background, showing a reduction by up to a factor 145 with pulse shape discrimination alone, is presented in this work.
\end{abstract}

\section{Introduction}

%The search for \bb{0} decay is a promising approach to search for physics beyond the Standard Model (SM). The decay would imply total lepton number violation and the Majorna nature of the neutrino. Furthermore, the absolute neutrino mass scale could be constrained with a positive measurement. 

The Germanium Detector Array (\gerda) experiment \cite{GERDA_NIM} located at the Laboratory Nazionali del Gran Sasso (Italy) of INFN is searching for the \bb{0} decay of \nuc{Ge}{76}. This isotope is embedded in HPGe detectors isotopically enriched to increase its fraction to about \unit[87]{\%}. An array of a few dozen HPGe detectors is immersed in a \unit[64]{m$^3$} tank filled wit LAr which serves as a cooling medium, an ultra pure passive radiation shield as well as an active scintillation veto in \PII\ of the experiment.
\gerda\ is proceeding in two phases: \PI\ was completed in May 2013 with a Ge exposure of \unit[21.6]{kg$\cdot$yr} and a background level of \unit[\baseTsolo{-2}]{cts/(keV$\cdot$kg$\cdot$yr)}. This resulted in a half-life limit of \unit[\baseT{2.1}{21}]{yr} (\unit[90]{\%} C.L) for the \bb{0} decay \cite{GERDA_0nubb}. The half-life of the SM allowed process of \bb{2} decay was measured with increased precision as \unit[\baseT{(1.926\pm 0.094)}{21}]{yr} \cite{GERDA_2nubb_v2}.
\PII\ of the experiment is currently (fall 2015) commissioning, aiming to increase the exposure to more than \unit[100]{kg$\cdot$yr} and to reduce the background level further by more than one order of magnitude. The backbone of the background reduction are 30 newly produced broad energy germanium (BEGe) detectors which have superior pulse shape characteristics compared to \PI\ semi-coaxial detectors \cite{GERDA_PSA}.

 The background analysis in \PI\ \cite{GERDA_bg} identified the major components of the background at \Qbb\ for BEGe detectors as \nuc{K}{42} on the n$^+$ detector surface (\unit[59]{\%}), \nuc{Bi}{214} and \nuc{Th}{228} in the detectors assembly (\unit[26]{\%}), \nuc{K}{42} homogeneously in the LAr (\unit[6]{\%}) and alpha emitters on the p$^+$ detector surface (\unit[3]{\%}) in addition to other smaller contributions.
The largest component of \nuc{K}{42} (\unit[Q$_\beta$=3525.4]{keV}, \unit[$T_{1/2}$=12.4]{h}) is part of the decay chain starting from \nuc{Ar}{42} (\unit[Q$_\beta$=599]{keV}, \unit[$T_{1/2}$=32.9]{h}) which is a trace isotope in natural argon. 
%with an abundance \baseT{<4.3}{-21} g/g (\unit[90]{\%} C.L.) \cite{Ashitkov03}. 
The high energetic electrons in the \nuc{K}{42} ground state transition (\unit[81.9]{\%} probability) can mimic the \bb{0} decay if it occurs close to the detector surface and if the electrons penetrate into the bulk detector volume. \nuc{K}{42} can be ionized after the \nuc{Ar}{42} decay and its attraction by the high voltage (HV) of the detectors was demonstrated. Additionally, \nuc{K}{42} concentrates on the n$^+$ detector surface \cite{GERDA_bg,GERDA_commissioning}.
%
%Additionally, \nuc{K}{42} can be ionized after the \nuc{Ar}{42} decay and its attraction by the high voltage (HV) of the detectors could be demonstrated. This results in a larger concentration on the n$^+$ detector surface \cite{GERDA_bg,GERDA_commissioning}. 
%
However, the electrons have to enter the detector volume passing through the \nPlus\ or the \pPlus\ which creates particular pulse shape features and enables a strong discrimination of these surface events. The understanding and simulation of the surface event discrimination will be shown in the following.

\section{Pulse Shape Discrimination with BEGe Detectors}

The \gerda\ BEGe detectors \cite{Canberra} are p-type HPGe detectors with an average mass of \unit[670]{g}. The \nPlus\ is created with Li doping and covers around \unit[98]{\%} of the detector surface with a thickness of \unit[0.5 to 0.9]{mm}. The \pPlus\ is created with B doping and has a thickness in the order of $\mu$m. 
When radiation creates electron-hole pairs, the charges are read out at the \pPlus\ with a charge sensitive pre-amplifier.
The weighting field strength inside the detector is mostly constant throughout the bulk and increases sharply around the \pPlus. 
For energy depositions inside the bulk, the holes drift first towards the detector center and then towards the \pPlus\ due to a specific configuration of space charges. This results in the same charge trajectory in the volume with large weighting field strength and thus in the same pulse shapes for practically all bulk events \cite{GERDA_PSA,Agostini11}.

%\begin{figure}[h]
%\begin{minipage}{17pc}
%\centering
%\includegraphics[width=12pc]{BEGeScheme_new.png}
%\caption{\label{BEGeScheme}BEGe detector}
%\end{minipage}\hspace{2pc}%
%\begin{minipage}{17pc}
%\centering
%\includegraphics[width=14.3pc]{GD91C_AOEmap.png}
%\caption{\label{GD91C_AOEmap}\AOE\ map inside BEGe detector}
%\end{minipage} 
%\end{figure}

\begin{figure}[b]
\includegraphics[width=25pc]{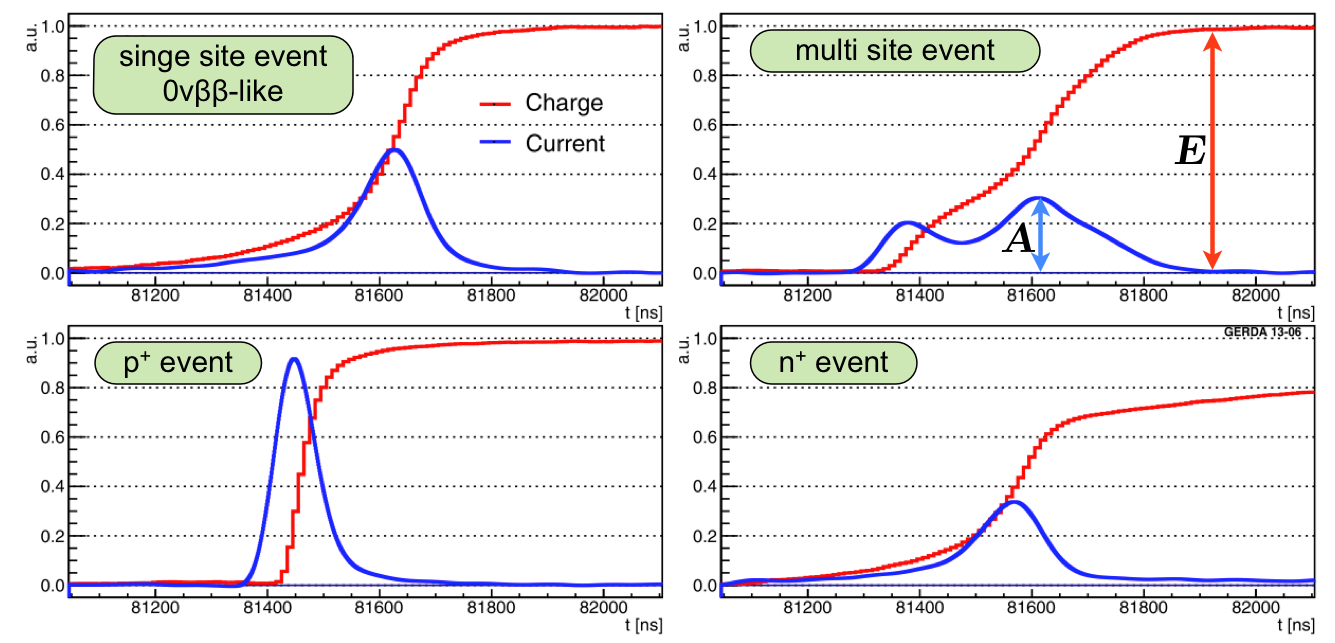}\hspace{2pc}%
\begin{minipage}[b]{11pc}\caption{\label{Interactions}Charge pulse shape (red) and current pulse shape (blue) for interactions in BEGe detectors. Multi site events created by $\gamma$ background, \pPlus\ events and \nPlus\ events created by surface background can be distinguished from bulk events created by \bb{0} decay. Taken from \cite{GERDA_PSA}}
\end{minipage}
\end{figure}

The experimental signature of the \bb{0} decay in \gerda\ is a discrete deposition of \unit[\Qbb=2039]{keV} by the two final state electrons in a small volume of the bulk of the detector. Thus the majority of \bb{0} events are single-site events (SSE) in the detector bulk which can be effectively distinguished from multi-site events (MSE) and surface events with the simple yet powerful pulse shape discriminator \AOE\ \cite{Budjas09}. The amplitude $A$ of the current pulse is obtained after differentiation of the charge signal and 3-fold shaping with a \unit[50]{ns} moving window average algorithm. After this soft shaping the current pulse still contains timing information of the charge collection. The energy $E$ is obtained after significantly stronger shaping \cite{GERDA_ZAC} which practically eliminates all timing information and results in an excellent energy resolution. The charge and current pulses of different event types are illustrated in \fig \ref{Interactions}. In case of MSE, multiple charge clusters arrive with a time offset due to different drift times. The amplitude and thus the \AOE\ is reduced for these events. In p$^+$ events the holes and electrons drift simultaneously inside the large weighting field and thus immediately induce a higher current. Those events have a larger amplitude and thus a larger \AOE\ value. In the \nPlus\ there is no electric field present and the dominant charge transport mechanism is diffusion. This results in two consequences: (1) The pulse formation is slower than for bulk events since the charges have to diffuse first into the region with electric field and (2) not all charges are collected resulting in a reduced energy measurement for interactions in a semi-active layer around the \nPlus.

The event characteristics of \nPlus\ interactions have been largely ignored in the past and the \nPlus\ was interpreted as a dead layer without charge collection. The dead layer interpretation, from now on called ``old model'' is valid for discrete energy depositions such as in gamma spectroscopy and \bb{0} decay search. However, the semi-active layer is effectively reducing the \nPlus\ thickness (aka dead layer thickness) for surface events and significantly changing their measured energy. In addition, the active volume for processes with continuous energy deposition such as \bb{2} decays is increased. The precise formation of those \nPlus\ ``slow pulses'' via diffusion is largely unknown and now studied with a ``new model'' of the \nPlus\ which can be folded into Geant4 Monte Carlo (MC) simulations via a post processing with pulse shape libraries. In the following, the new model is described, verified with calibration source measurements and applied to \nuc{K}{42} surface events and \bb{2} decays in \gerda\ BEGe detectors.

\section{New Model for n$^+$ Surface Events}

%\begin{figure}[h]
%\includegraphics[width=20pc]{pdf_LiProfileFCCD.png}\hspace{2pc}%
%\begin{minipage}[b]{15pc}\caption{\label{pdf_LiProfileFCCD}Construction of the n$^+$ model from first principles. The Li concentration profile is calculated with diffusion. The three regions of RDR, DDR and FAV are determined according to the Li concentration and the FCCD measurement. The n$^+$ model is constructed individually for each detector. Charges are propagated accordingly: RDR - diffusion and recombination, DDR - diffusion, FAV - drift in electric field and full collection.}
%\end{minipage}
%\end{figure}

\begin{figure}[h]
\begin{minipage}{16pc}
\centering
\includegraphics[width=16pc]{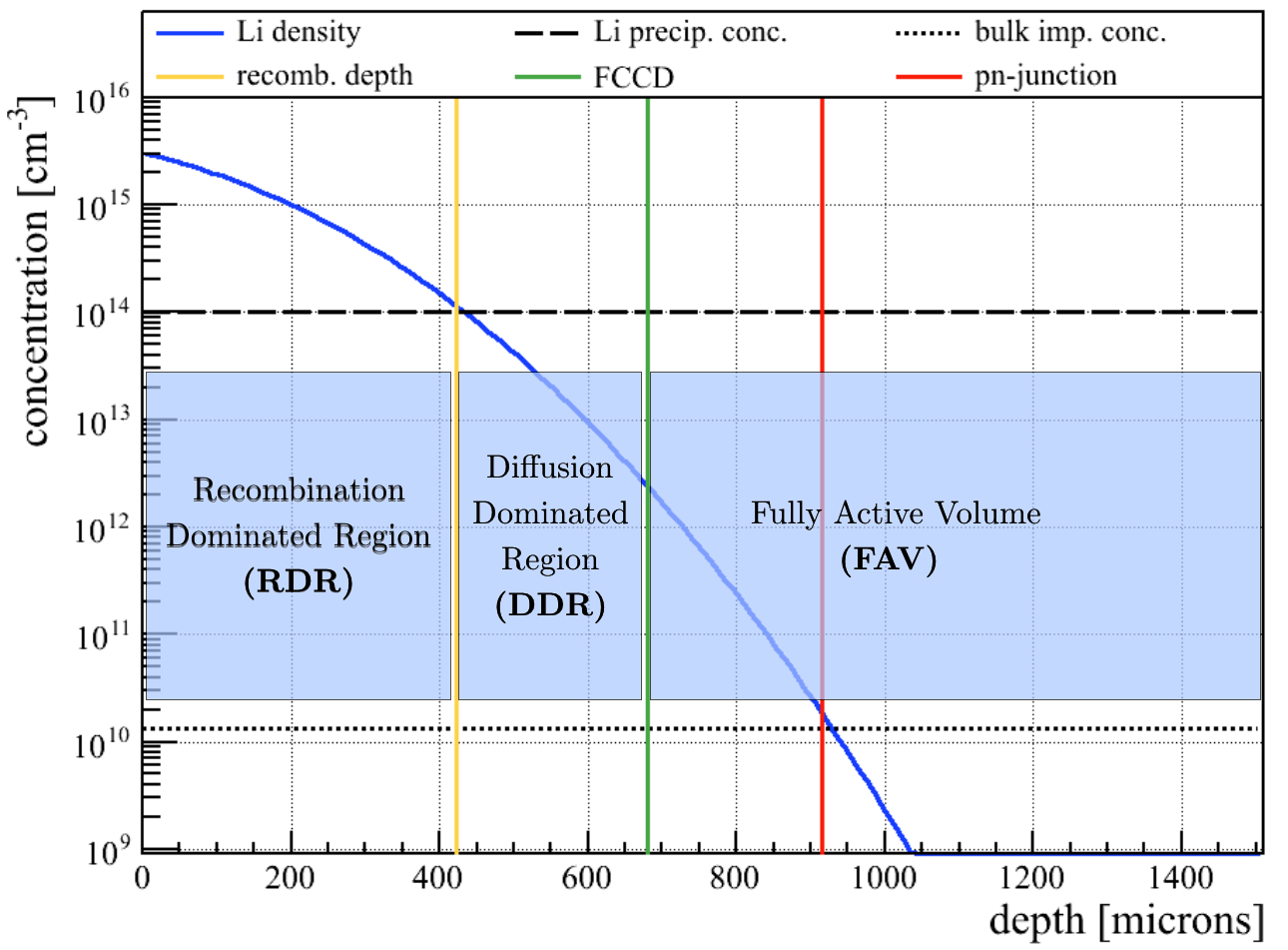}
\caption{\label{pdf_LiProfileFCCD}In the model the n$^+$ electrode is separated into three regions according to the Li concentration profile.}
\end{minipage}\hspace{2pc}%
\begin{minipage}{18pc}
\centering
\includegraphics[width=18pc]{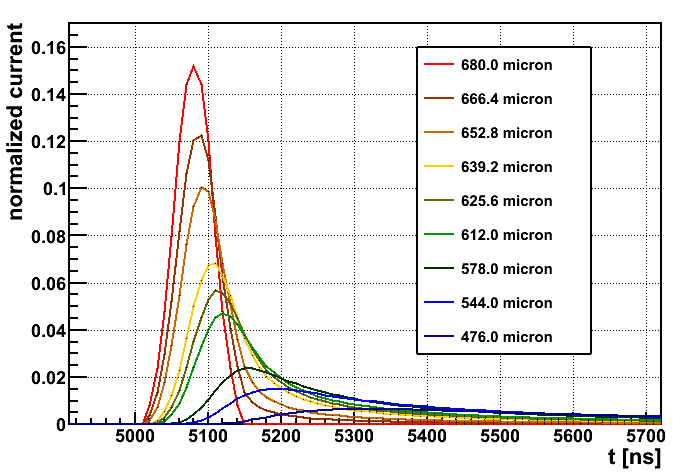}
\caption{\label{pdf_CurrentPulsesMWA}Simulated current pulses for different interaction depths in the \nPlus\ of a typical detector with \unit[680]{micron} FCCD.}
\end{minipage} 
\end{figure}

The \nPlus\ of \gerda\ BEGe detectors is created with Li diffusion during a \unit[2-4]{h} annealing cycle at \unit[200]{$^{\circ}$C}. The solution of the diffusion equation assuming a constant concentration on the surface is shown as the blue line in \fig \ref{pdf_LiProfileFCCD}.
The n$^+$ layer is divided into three regions according to the Li concentration profile \cite{PhDFinnerty}: (1) The fully active volume (FAV) in which charges drift in the electric field and are completely collected, (2) the drift dominated region (DDR) where no electric field is present and charges diffuse until they either reach the FAV or they reach the (3) recombination dominated region (RDR) where the charges recombine with a certain probability over time. The boundaries of these regions are constructed from first principles and measurements and are individual for each BEGe detector.

The p-n junction depth is defined where the Li donor concentration is equal to the bulk acceptor concentration. When the detector is biased and its bulk volume is fully depleted of free charge carriers, the number of charges in the bulk region compensate the same number of charges in the \nPlus\ creating a depletion zone which is equivalent to the FAV. 
The border of the depletion zone defines the full charge collection depth (FCCD). Due to the steep Li concentration profile, the FCCD is only \order{\unit[100]{$\mu$m}} above the p-n junction. After the annealing at high temperature, the Li concentration is supersaturated at room and LAr temperatures in the outer layers of the \nPlus. The Li forms precipitation centers which are electrically inert and believed to be responsible for charge recombination \cite{PhDFinnerty}. The recombination depth defined as the boundary between DDR and RDR is taken at a Li concentration of \unit[\baseTsolo{14}]{cm$^{-3}$} (see Ref.~\cite{PhDLehnert} for details). The FCCD is measured for each \gerda\ detector \cite{GERDA_BEGeTests,PhDLehnert} and used to recursively determine the Li concentration profile and the other region boundaries.

In these regions charges are simulated with the diffusion equation in a \unit[1]{ns} time and \unit[10]{$\mu$m} spatial dimension. 
%A simulation with three spatial dimensions did not show any significant difference to the results. 
Charges passing the FCCD entering the FAV are collected and form the current pulse. Charges passing into the RDR recombine with a probability of \unit[0.002]{ns$^{-1}$}. This value was determined when comparing different recombination rates with calibration data \cite{PhDLehnert}. A model with zero recombination rate as well as a model with immediate recombination is strongly disfavoured by the data.
Charges starting at a given depth below the surface result in very different pulse shapes which are collected in a pulse shape library depending on the interaction depth. Example pulses are shown in \fig \ref{pdf_CurrentPulsesMWA}. The amplitude and the integral charge are reduced the more shallow the interaction.
This surface interaction pulse shape library is combined with a bulk pulse shape library created with a modified version of ADL \cite{PhDMarco}. 
%The \AOE\ values of all locations in the detector are shown in \fig \ref{GD91C_AOEmap}. 
Geant4 simulations with the MaGe framework \cite{MaGe} are post processed with these two libraries such that the \AOE\ information is extracted for an event based on the superposition of the individual pulse traces for each Geant4 hit in the event \cite{PhDLehnert}.
%
%
%\begin{figure}[h]
%\begin{minipage}{16pc}
%\includegraphics[width=16pc]{pdf_CurrentPulsesMWA.png}
%\caption{\label{pdf_CurrentPulsesMWA}Figure caption for first of two sided figures.}
%\end{minipage}\hspace{2pc}%
%\begin{minipage}{16pc}
%\includegraphics[width=16pc]{pdf_CCE.png}
%\caption{\label{pdf_CCE}Figure caption for second of two sided figures.}
%\end{minipage} 
%\end{figure}
%
%
%
The MC simulations with the post processing of the new n$^+$ model are validated with \nuc{Th}{228}, \nuc{Am}{241}, \nuc{Sr}{90} calibration source measurements. The low \AOE\ values of \nuc{Tl}{208} \glines\ and the single escape peak at \unit[2103.5]{keV} are almost exclusively created by MSE effects. On the other hand, the low \AOE\ values of the \nuc{Am}{241} \unit[59.5]{keV} \gray\ are almost exclusively created by the \nPlus. The low \AOE\ events of the \nuc{Sr}{90} source are created by betas being influenced by the \nPlus\ and by MSE effects from hard Bremsstrahlung. This beta source is taken as a proxy for \nuc{K}{42} decays in \gerda. The data of these calibration sources with different event topologies is well described by the MC simulations. For details see Ref.~\cite{PhDLehnert}.

%The \AOE\ value versus the energy is plotted for events from a measured and simulated \nuc{Sr}{90} calibration source (\unit[2280]{keV} beta endpoint of \nuc{Y}{90}) in \fig \ref{pdf_Sr90_AOE_poster_EXP_UDB} and \fig \ref{SpecSr90_Scatter_KR2}, respectively. SSE in the bulk are normalized to \AOE=1 and the exposure of the detectors surface is illustrated in \fig \ref{GD91C_AOEmap}. The following features can observed: SSE events in the bulk dominate at lower energies. These events are created by Bremsstrahlung \grays\ which interact in the detector bulk. At higher energies the events populate a band at $\rm A/E<1$. Electrons passing through the \nPlus\ are continously ionizing and always have a fraction of there pulse influenced by the slow pulse character. Additionally the low \AOE\ region is also populated by MSE events created by harder Bremsstrahlung \grays. The population of events at $\rm A/E>1$ is due to scatted electrons interacting around the \pPlus. The possibility to separate n$^+$ and p$^+$ surface events depends on the gap between the signal region at $\rm A/E=1$ and the populations at $\rm 1<A/E>1$ which is very dependent on the specific detector and its FCCD. For the example shown in \fig \ref{pdf_Sr90_AOE_poster_EXP_UDB} a \nuc{Sr}{90} suppression of \missing can be reached with an \AOE\ cut of $\rm 0.98<A/E<1.07$. The suppression obtained from the new model can reproducing this strong suppression, however conservatively underestimates it by a factor of \missing.\\  
%%

\section{Application to Signal and Background Processes in GERDA Phase II}

\begin{figure}[b]
\begin{minipage}{16pc}
\includegraphics[width=16pc]{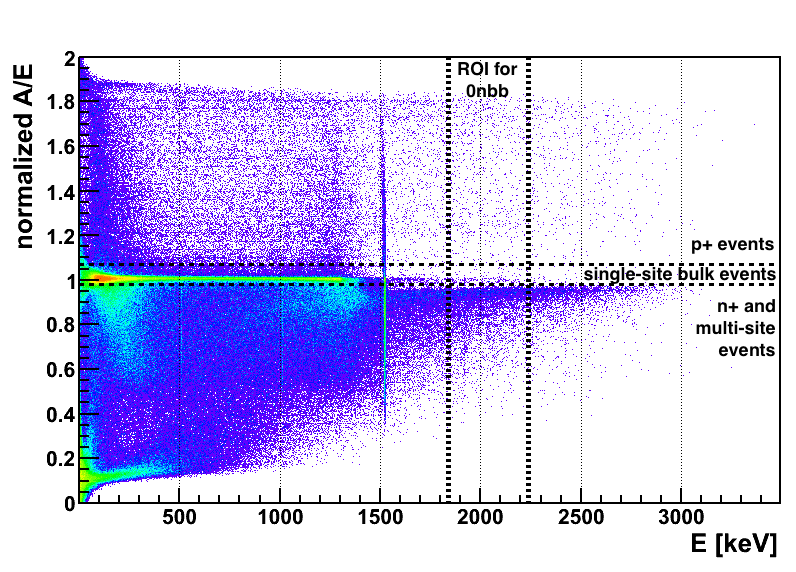}
\caption{\label{pdf_AOEvsE_K42}\AOE\ versus energy of simulated \nuc{K}{42} events in LAr around a BEGe detector.}
\end{minipage}\hspace{2pc}%
\begin{minipage}{16pc}
\includegraphics[width=16pc]{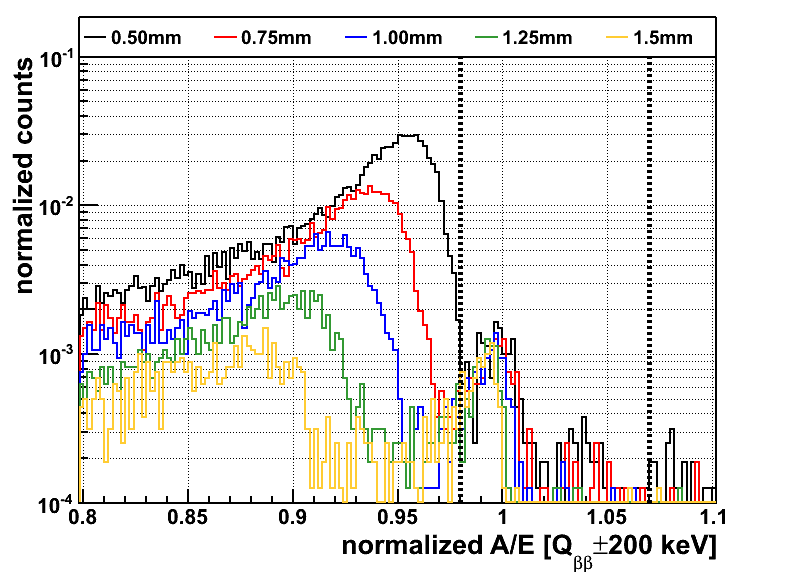}
\caption{\label{pdf_AOEQbb_K42_CompFuture}\AOE\ in \unit[\Qbb$\pm200$]{keV} for simulated \nuc{K}{42} events around BEGe detectors with different FCCD.}
\end{minipage} 
\end{figure}

\begin{figure}[h]
\begin{minipage}{16pc}
\includegraphics[width=16pc]{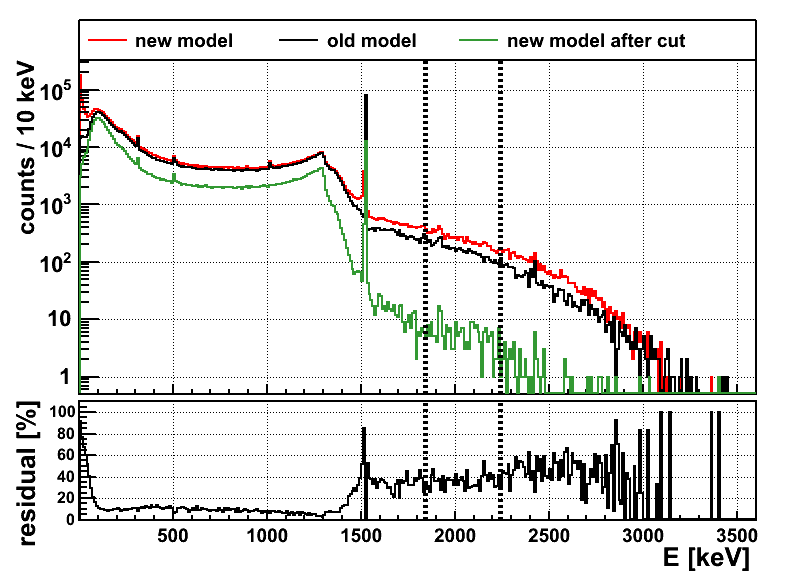}
\caption{\label{pdf_ESpecRes_K42}Spectrum of simulated \nuc{K}{42} events for the new model, old model, and the new model after \AOE\ cut.}
\end{minipage}\hspace{2pc}%
\begin{minipage}{16pc}
\includegraphics[width=16pc]{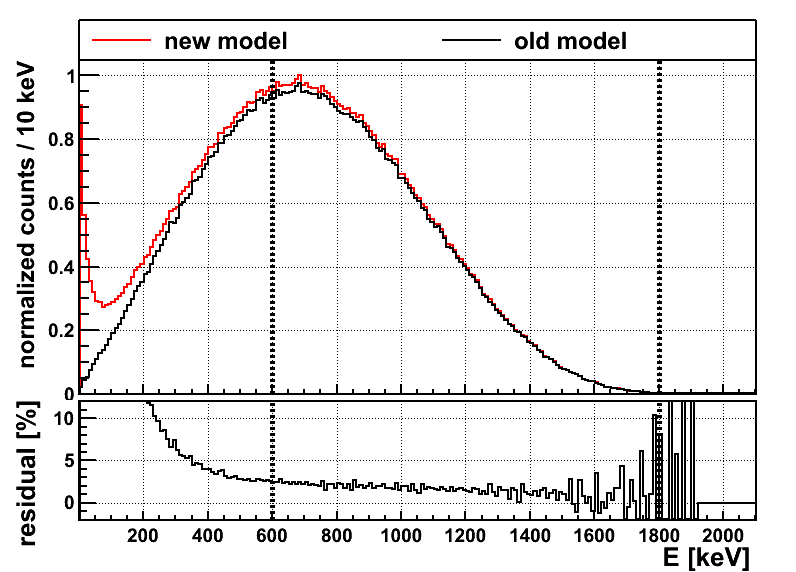}
\caption{\label{pdf_ESpecRes_2nbb}Energy spectrum of simulated \bb{2} decays for new and old model.}
\end{minipage} 
\end{figure}

The new model is applied to simulations of \nuc{K}{42} decays in the LAr around a typical \gerda\ detector with \unit[0.68]{mm} FCCD. The \AOE\ versus energy spectrum is shown in \fig \ref{pdf_AOEvsE_K42} in which SSE in the bulk are normalized to $A/E=1$. The following features can be seen: The \gline\ at \unit[1525]{keV} (\unit[18]{\%} probability) has high and low \AOE\ values coming from interactions close to the \pPlus\ and MSE interactions, respectively. The Compton region below the \gline\ is dominated by SSE interactions. The region above the peak, especially in \unit[\Qbb$\pm$]{200 keV} illustrated with vertical dashed lines, is dominated by interactions of electrons passing through the \nPlus. A part of their energy is deposited in the semi-active layer thus shifting the \AOE\ for this population collectively into a band below the SSE bulk value of $A/E=1$. 
The \AOE\ values for these events are shown in \fig \ref{pdf_AOEQbb_K42_CompFuture} for various detectors with different FCCD. The gap between the SSE band and the slow pulse band and thus the discrimination possibility for n$^+$ surface events is larger for detectors with thicker \nPlus. Surviving events in the SSE region are due to Bremsstrahlung \grays\ jumping the \nPlus\ and depositing energy solely in the bulk. In this energy region a \nuc{K}{42} suppression of 45 can be reached with a cut of $\rm 0.98<A/E<1.07$. The suppressed energy spectrum of \nuc{K}{42} is shown in \fig \ref{pdf_ESpecRes_K42} (green) compared to the unsuppressed spectrum (red). The semi-active layer increases the number of events around \Qbb\ by \unit[35]{\%} which can be seen in the comparison with the old model (black) in the residual plot.
The suppression is further increased by the LAr scintillation veto by more than a factor of 3 for BEGe detectors. 
For \nuc{K}{42} events directly on the \nPlus\ there is less chance to create a Bremsstrahlung \gray\ without also depositing energy in the \nPlus\ and consequently reducing the \AOE\ value of the event. The suppression in this case is up to a factor 145 with the same \AOE\ cut. However, for these events the LAr veto is not effective since the beta is directly entering the detector without depositing energy in the LAr \cite{PhDLehnert}. 

For the \bb{2} decay the active volume is increases when part of the \nPlus\ is semi-active compared to the old dead layer assumption. Applying the new model to \bb{2} decays inside a typical detector results in a \unit[2]{\%} increase in active volume in the energy range of \unit[$600-1800$]{keV} \cite{PhDLehnert}. In addition the spectral shape changes with the largest effects below \unit[600]{keV} (\fig \ref{pdf_ESpecRes_2nbb}). This region is however not accessible in \gerda\ due to the dominating background of \nuc{Ar}{39} (\unit[565]{keV} beta endpoint). The new n$^+$ model will help to increase the precision of future \nuc{Ge}{76} \bb{2} decay half-life measurements which have a current precision of \unit[5]{\%} \cite{GERDA_2nubb_v2}.

\section{Conclusions}

A new \nPlus\ model was developed from first principles based on diffusion of the charge carriers. This model was folded into Geant4 MC simulations and tuned and compared to calibration source measurements resulting in a good description of the data. The new model was applied to simulations of the dominating \nuc{K}{42} surface background predicting a suppression factor of up to $145$. This is the first time that the large suppression of surface backgrounds is understood in detail.
Additionally, the active volume and the spectral shape of \bb{2} decays is affected by the proper treatment of the \nPlus. 
The effective volume for \bb{2} is increased by \unit[2]{\%} when including the semi-active layer, which reduces the \bb{2} decay rate inferred from a measurement.

\end{document}

%% file: commands.tex
\newcommand{\nuc}[2]{$^{#2}\rm #1$}

\newcommand{\bb}[1]{$\rm #1\nu \beta \beta$}
\newcommand{\bbm}[1]{$\rm #1\nu \beta^- \beta^-$}
\newcommand{\bbp}[1]{$\rm #1\nu \beta^+ \beta^+$}
\newcommand{\bbe}[1]{$\rm #1\nu \rm ECEC$}
\newcommand{\bbep}[1]{$\rm #1\nu \rm EC \beta^+$}

\newcommand{\rootcern}{\textsc{Root}}
\newcommand{\gerda}{\textsc{Gerda}}
\newcommand{\largeGERDA}{{LArGe}}
\newcommand{\PI}{\mbox{Phase\,I}}
\newcommand{\PIa}{\mbox{Phase\,Ia}}
\newcommand{\PIb}{\mbox{Phase\,Ib}}
\newcommand{\PIc}{\mbox{Phase\,Ic}}
\newcommand{\PII}{\mbox{Phase\,II}}

\newcommand{\geant}{\textsc{Geant4}}
\newcommand{\mage}{\myacs{MaGe}}
\newcommand{\decayzero}{\textsc{Decay0}}

\newcommand{\nPlus}{\mbox{n$^+$ electrode}}
\newcommand{\pPlus}{\mbox{p$^+$ electrode}}

\newcommand{\AOE}{$A/E$}

\newcommand{\order}[1]{\mbox{$\mathcal{O}$(#1)}}

\newcommand{\mul}[1]{\texttt{multiplicity==#1}}

\newcommand{\pic}[5]{
       \begin{figure}[ht]
       \begin{center}
       \includegraphics[width=#2\textwidth, keepaspectratio, #3]{#1}
       \caption{#5}
       \label{#4}
       \end{center}
       \end{figure}
}

\newcommand{\apic}[5]{
       \begin{figure}[H]
       \begin{center}
       \includegraphics[width=#2\textwidth, keepaspectratio, #3]{#1}
       \caption{#5}
       \label{#4}
       \end{center}
       \end{figure}
}

\newcommand{\sapic}[5]{
       \begin{figure}[P]
       \begin{center}
       \includegraphics[width=#2\textwidth, keepaspectratio, #3]{#1}
       \caption{#5}
       \label{#4}
       \end{center}
       \end{figure}
}

\newcommand{\picwrap}[9]{
       \begin{wrapfigure}{#5}{#6}
       \vspace{#7}
       \begin{center}
       \includegraphics[width=#2\textwidth, keepaspectratio, #3]{#1}
       \caption{#9}
       \label{#4}
       \end{center}
       \vspace{#8}
       \end{wrapfigure}
}

\newcommand{\baseT}[2]{\mbox{$#1\cdot10^{#2}$}}
\newcommand{\baseTsolo}[1]{$10^{#1}$}
\newcommand{\THL}{$T_{\nicefrac{1}{2}}$}

\newcommand{\UBI}{$\rm cts/(kg \cdot yr \cdot keV)$}

\newcommand{\Uflux}{$\rm m^{-2} s^{-1}$}
\newcommand{\Ucpd}{$\rm cts/(kg \cdot d)$}
\newcommand{\Uexpo}{$\rm kg \cdot d$}
\newcommand{\UexpoYear}{$\rm kg \cdot yr$}

\newcommand{\UMWE}{m.w.e.}

\newcommand{\Qbb}{$Q_{\beta\beta}$}

\newcommand{\validate}{\textcolor{blue}{\textit{(validate!!!)}}}

\newcommand{\improve}{\textcolor{blue}{\textit{(improve!!!)}}}

\newcommand{\missing}{\textcolor{red}{\textbf{...!!!...} }}

\newcommand{\quanta}{\textcolor{red}{\textit{(quantitativ?) }}}

\newcommand{\misscite}{\textcolor{red}{[citation!!!]}}

\newcommand{\missref}{\textcolor{red}{[reference!!!]}\ }

%K42
\newcommand{\PC}{$N_{\rm peak}$}
\newcommand{\BIC}{$N_{\rm BI}$}
\newcommand{\PAPR}{$R_{\rm p/>p}$}

\newcommand{\PCR}{$R_{\rm peak}$}

%Pd

\newcommand{\gline}{$\gamma$-line}
\newcommand{\glines}{$\gamma$-lines}

\newcommand{\gray}{$\gamma$-ray}
\newcommand{\grays}{$\gamma$-rays}

\newcommand{\bray}{$\beta$-ray}
\newcommand{\brays}{$\beta$-rays}

\newcommand{\aray}{$\alpha$-ray}
\newcommand{\arays}{$\alpha$-rays}

\newcommand{\betas}{$\beta$'s}

%general

\newcommand{\tab}{{Tab.~}}
\newcommand{\eq}{{Eq.~}}
\newcommand{\fig}{{Fig.~}}
\renewcommand{\sec}{{Sec.~}}
\newcommand{\chap}{{Chap.~}}

 \newcommand{\fn}{\iffalse \fi} %footnote explaination
 \newcommand{\tx}{\iffalse \fi} %text explaination
 \newcommand{\txe}{\iffalse \fi} %text extended explaination
 \newcommand{\sr}{\iffalse \fi} %section reference explaination